\documentclass[aps,twocolumn,showpacs,pra,groupedaddress,superscriptaddress,nofootinbib]{revtex4}

\usepackage{graphicx,amsmath,amssymb,amsbsy,subfigure,hyperref,times}

\hypersetup{
   colorlinks=true,
   linkcolor=blue,
}
\graphicspath{{figure/}}
\begin{document}

\newcommand{\ket}[1]{|#1\rangle}
\newcommand{\bra}[1]{\langle#1|}
\newcommand{\ketbra}[1]{| #1\rangle\!\langle #1 |}
\newcommand{\kebra}[2]{| #1\rangle\!\langle #2 |}
\newcommand{\id}{\mathbbm{1}}
\newcommand{\ohm}{\Omega_{\rm CQ}}
\newcommand{\rhobd}{\rho^{\vec{c}}_{AB}}

\providecommand{\tr}[1]{\text{tr}\left[#1\right]}
\providecommand{\tra}[1]{\text{tr}_A\left[#1\right]}
\providecommand{\trb}[1]{\text{tr}_B\left[#1\right]}
\providecommand{\abs}[1]{\left|#1\right|}
\providecommand{\sprod}[2]{\langle#1|#2\rangle}
\providecommand{\expect}[2]{\bra{#2} #1 \ket{#2}}


\title{Dynamics of Quantum Coherence and Quantum Fisher Information of a V-type Atom in Isotropic Photonic Crystal}
\author{Ghafar Ahmadi}
\affiliation{Department of Physics, Institute for Advanced Studies in Basic Sciences (IASBS), Zanjan, Iran.}
\author{Shahpoor Saeidian}
\email{saeidian@iasbs.ac.ir}
\affiliation{Department of Physics, Institute for Advanced Studies in Basic Sciences (IASBS), Zanjan, Iran.}
\author{Ghasem Naeimi}
\email{ghnaeimi@iau.ac.ir}
\affiliation{Department of Physics, Qazvin Branch, Islamic Azad University, Qazvin, Iran}

\begin{abstract}
The time evolution of quantum Fisher information, quantum coherence, and non-Markovianity of a V-type three-level atom, considering all spontaneous emissions, both in free space and inside a photonic band gap crystal, are investigated. It has been demonstrated that the photonic band gap crystal, as a structured environment, significantly influences the preservation and enhancement of these quantum features. Additionally, we observe that by manipulating the initial relative phase values encoded in the atomic state and the relative positions of the upper levels within the forbidden gap, control over the dynamics of quantum features can be achieved.
These findings highlight the potential benefits of utilizing photonic band gap crystals in quantum systems, offering improved preservation and manipulation of quantum information. The ability to control quantum features opens new avenues for applications in quantum information processing and related technologies.
\end{abstract}

\date{\today}


\maketitle

\section{Introduction}

In the field of quantum mechanics, estimation plays a crucial role in extracting valuable information from quantum systems. One of the fundamental challenges in quantum physics is accurately determining the values of unknown parameters that characterize a quantum system, such as the precise value of a physical quantity or the state of a quantum system.  Parameter estimation is essential in quantum metrology \cite{Dowling, pirandolaReview, Zhang2013, demkowicz2015quantum, Najafi, ALBARELLI2020126311,Naeimi1}. Quantum metrology aims to utilize of quantum features and effects to enhance the sensitivity in estimating physical parameters. Quantum Estimation Theory (QET) has emerged as a powerful framework to address the problem of parameter estimation in quantum systems \cite{helstrom1969quantum, suzuki2019information, paris2009quantum, Metro22}. QET provides a mathematical formalism to analyze and optimize measurement schemes for extracting unknown parameters with high precision.  At the core of QET is the concept of Quantum Fisher Information (QFI), which quantifies the information content contained in a given quantum state and determines the fundamental limits of parameter estimation.  A larger value of QFI indicates a higher precision in parameter estimation. QFI has been widely investigated theoretically and experimentally for different systems \cite{szczykulska2016multi, mitchell2004super, Knott, REN2022105542, Zhang2013, Yu2021, yousefi2022quantum,ALGARNI2022106089,Ren2018}.
A key factor that affects estimation in quantum systems is the interaction between the open quantum system and its surrounding environment \cite{ref30, lidar2020lecture, Anwar2023}. This interaction can lead to decoherence, where the fragile quantum states lose their coherence and become entangled with the environment. Decoherence causes a loss of information and degrades the accuracy of parameter estimation. Therefore, understanding the effects of the environment and mitigating decoherence is vital for precise estimation in quantum mechanics.
One way to characterize the effects of the environment on quantum coherence and parameter estimation is through the distinction between Markovian and non-Markovian dynamics. In a Markovian environment, the memory of past interactions between the quantum system and the environment is lost, and the dynamics are memoryless. This simplifies the analysis but limits the control over the environment's effects on the quantum system. On the other hand, non-Markovian dynamics in open quantum systems exhibit memory effects. The system's evolution depends not only on its current state but also on its past history, leading to time-dependent dynamics with memory, revivals, or oscillations \cite {LI20181, BreuerPhysRevLett, LuoPhysRevA, Jahromi, Mahdavipour, yousefi2022quantum}.
By using structured environments or engineering the system-environment interactions, non-Markovian dynamics can be utilized to enhance the accuracy and robustness of estimation in quantum mechanics \cite {Bennink_2019, ChinPhysRevLett, escher2011general, alipour2014quantum, xiao2014distribution}.
One fascinating approach to controlling the effects of the environment on quantum systems is through the concept of photonic band gap (PBG) crystal. PBG refer to ranges of frequencies or wavelengths in which the propagation of light is forbidden within a material or a photonic crystal structure.  In such structures, the interference of waves creates regions where specific wavelengths of light cannot propagate, leading to the confinement of light within certain regions \cite{JohnPhysRevLett, woldeyohannes1999coherent, Singh}.
By engineering PBG, researchers can create environments that influence the interaction of light with quantum systems.  PBG materials can be utilized to control the propagation and emission of photons, which are essential for various quantum technologies such as waveguides \cite{d2003photonic, kaniber2007efficient}, optical memory devices \cite{spraque2014broadband, quang1997coherent}, and single-photon sources \cite{kaniber2008highly, laucht2012waveguide}. Atom–photon entanglement \cite{ entezar2010controllable} and the control of the spontaneous emission of atoms within photonic band gap materials have been investigated \cite{woldeyohannes2019optical,Mesfin2003, jiang2014control,yang2000sontaneous,huang201calculation}. Recently, it has been shown that PBG materials can help to confine and protect quantum states, minimize decoherence effects and enhancing the precision of parameter estimation \cite{ yousefi2022quantum}. They have studied the dynamics of the quantum features of a three-level atom coupled to a classical field by considering the leading approximation. In their work aiming to reach an analytical solution, the leading approximation has been utilized wherein a number of spontaneous emission effects and non-radiative interactions have been neglected. In our paper, we study the time evolution of quantum Fisher information, quantum coherence, and non-Markovianity of a V-type three-level atom embedded in free space or in a PBG crystal. Notably, classical field is not required, and all spontaneous emissions are accounted for to achieve the analytical solution. We observed significant differences in the characteristics of quantum Fisher information, quantum coherence, and non-Markovianity between free space and a PBG crystal, stemming from environmental distinctions. We assess the influence of the initial relative phase values encoded in the atomic state and the relative positions of the upper levels within the forbidden gap over the dynamics of quantum features. Our results demonstrate that by properly choosing the values of these two parameters, we can effectively control the dynamics of quantum features.

The structure of the paper is as follows: In Sec. II, we describe the model system and obtain the analytical expression for probability amplitudes of the atomic system when the atom embedded in PBG and free space. In Sec. III, The dynamical behavior of quantum Fisher information and parameter estimation for both cases is studied. In Sec. IV, we discuss the dynamical behavior of quantum coherence for both situation. In Sec. V, we also delve into the dynamical behavior of non-Markovianity based on Hilbert Schmidt Speed (HSS). Finally, in Sec. VI, we provide a short conclusion.

\section{Model and Equations}\label{secBell}

\begin{figure}[t!]
\includegraphics[width=0.3\textwidth]{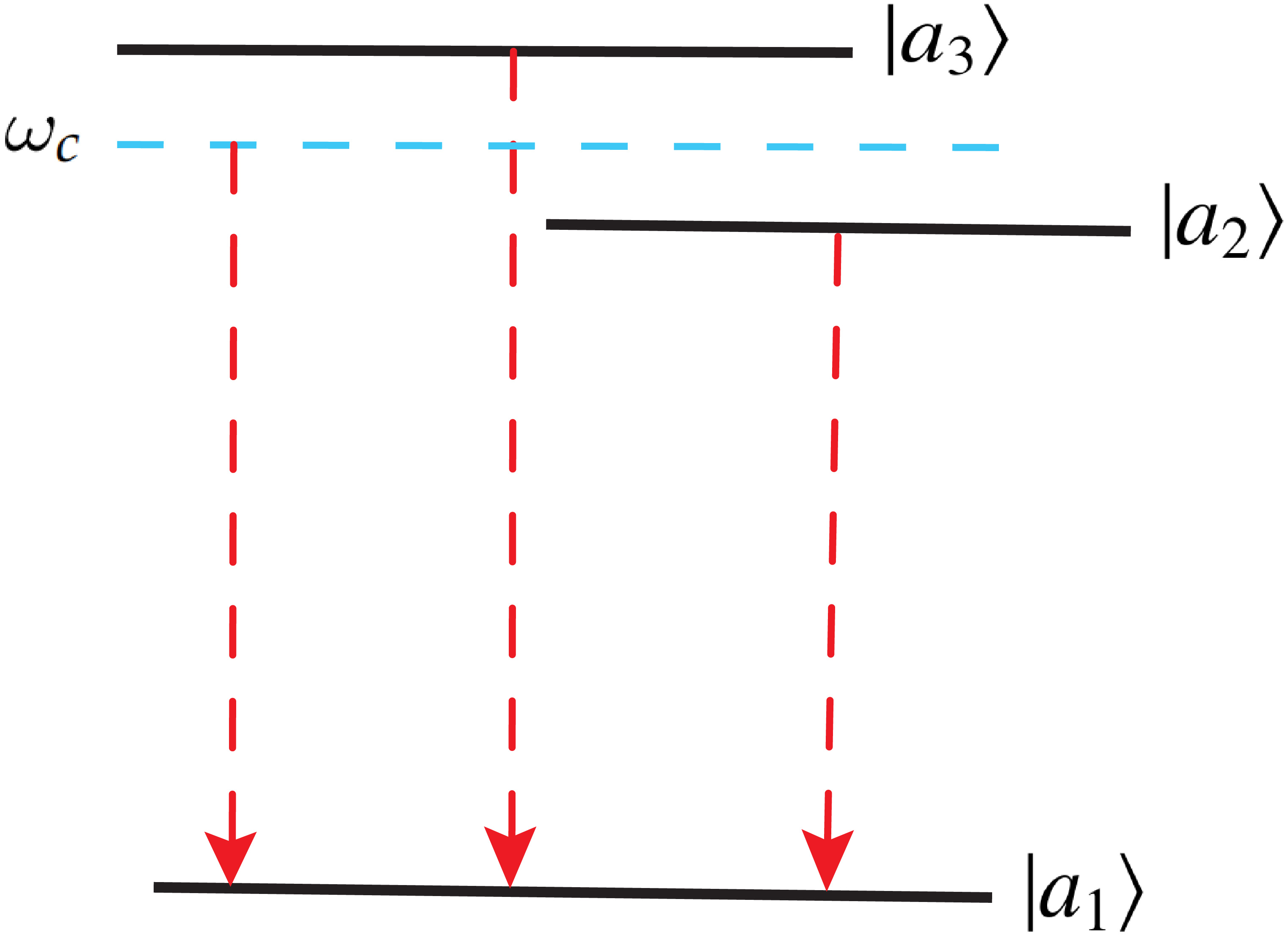}
\caption{Schematic diagram of a three-level atomic system in the V configuration}
\label{Fig.1}
\end{figure}

We consider a V-type three-level atom with two upper levels $\ket{a_3}$, $\ket{a_2}$, and a lower level $\ket{a_1}$ as shown in Fig.\ref{Fig.1}. This atomic scheme can be experimentally realized using  a $^{87}\mathrm{Rb}$  atom with $5^{2}S_{1/2} (F = 2)$ , $5^{2}P_{1/2}(F = 1)$, and  $5^{2}P_{3/2}(F = 3)$ representing the $\ket{a_1}$, $\ket{a_2}$, and  $\ket{a_3}$  states respectively  \cite{Steck, Higgins2021ElectromagneticallyIT,DeRose2023}.
The upper levels  $\ket{a_3}$,  $\ket{a_2}$ are coupled by the same vacuum mode to the lower level $\ket{a_1}$, and the transition between upper  levels $\ket{a_3}$, $\ket{a_2}$ is forbidden due to symmetry consideration. 
The transition frequencies between exited state $\ket{a_3}$, $\ket{a_2}$ and the ground state $\ket{a_1}$ are $\omega_{31}=\omega_{3}-\omega_{1}$ and $\omega_{21}=\omega_{2}-\omega_{1}$, respectively.
The Hamiltonian of this system with the rotating-wave approximation and the electric-dipole approximation can be written as

\begin{equation}
\label{eq:1}
\hat{H}=\hat{H_0} +\hat{H_I},
\end{equation}
where
\begin{equation}
\label{eq:2}
\hat{H_0} = \hslash \omega_{3}\ket{a_3}\bra{a_3}+\hslash \omega_{2}\ket{a_2}\bra{a_2}+\sum_{\lambda=1}^{2}\sum_{k}^{} \hslash \omega_{k}\hat{a}_{k \lambda}^{\dagger}\hat{a}_{k \lambda},
\end{equation}

\begin{eqnarray}\label{eq:3}
	\hat{H_I} =i \hslash\sum_{k \lambda}\left[ g_{k,\lambda}^{(31)} \hat{a}_{k \lambda}^{\dagger}\ket{a_1}\bra{a_3} 
	+g_{k,\lambda}^{(21)} \hat{a}_{k \lambda}^{\dagger}\ket{a_1}\bra{a_2}\right]+H.C.
\end{eqnarray}
Where $\hat{a}_{k \lambda}$ and $\hat{a}_{k \lambda}^{\dagger}$   are the annihilation and creation operators for the $k^{th}$ electromagnetic mode with frequency $\omega_{k}$ , and $\hbar k$ and $\lambda$ indicate momentum and two transverse polarization of electromagnetic mode, respectively.
The coupling constants between the $k^{th}$ electromagnetic mode and the atomic transitions $\ket{a_i}\hspace{1mm} (i= 2,3)$ are

\begin{equation}
	\label{eq:4}
	g_{k,\lambda}^{(i1)}= \frac{\omega_{i} d_{i}}{\hslash}\left(\frac{\hslash}{2\epsilon_{0}\omega_{k}V}\right)^{1/2} \hat{e}_\mathrm{k \lambda}.\hat{u}_\mathrm{i},
\end{equation}
that are assumed to be real. $d_{i}$ and $\hat{u_i}_\mathrm{}$ are the magnitude and unit vector of transition dipole moment of the atom, $V$ is the sample volume , $\hat{e}_{k,\lambda}$ are the two transverse unit vectors, and $\epsilon_{0}$ is the vacuum permittivity.
We assume that the photon reservoir is initially in the vacuum state $\ket{0}$ and the atomic system is prepared in a pure superposition of the two upper levels $\ket{a_3}$ and $\ket{a_2}$.  Therefore, for the initial vector state of the system we have

\begin{equation}
	\label{eq:5}
    \ket{\psi(0)}=\cos\left(\frac{\theta}{2}\right) \ket{a_3,0}+e^{i\phi} \sin\left(\frac{\theta}{2}\right)\ket{a_2,0},
\end{equation}
the  parameter $\theta$ measures the degree of  superposition of the two upper levels $\ket{a_3}$, $\ket{a_2}$, and the factor $e^{i\phi}$ gives the relative phase between the expansion coefficients of $\ket{a_3}$ and $\ket{a_2}$. The state vector of the system at arbitrary time t can be written as

\begin{eqnarray}
	\label{eq:6}
	\nonumber
	\ket{\psi(t)}&=&A_{3}(t) e^{-i\omega_{3}t} \ket{a_3,0}+A_{2}(t) e^{-i\omega_{2}t}\ket{a_2,0} \\ 
	&+& \sum_{k \lambda} A_{k \lambda}(t)e^{-i\omega_{k}t} \ket{a_1,1_{k \lambda}},
\end{eqnarray}
Above, the state vectors  $\ket{a_3,0}$ and $\ket{a_2,0}$ indicate the atom in its excited states $\ket{a_3}$ and $\ket{a_2}$ respectively, without any photons in the reservoir modes, and the state vector $\ket{a_1,1_{k \lambda}}$  indicates the atom in its ground state $\ket{a_1}$, with a single photon in the $k^{th}$ electromagnetic mode. 

Substituting the Hamiltonian (\ref{eq:1}) and the state vector (\ref{eq:6}) into the Schrödinger equation, we can obtain the  coupled  equations for the amplitudes as follows:

\begin{subequations} \label{eq:subeqns}
	\begin{equation}
		\label{eq:subeq1}
		{\dot{A}_{3}(t)}  = -\sum_{k \lambda} g_{k,\lambda}^{(31)} A_{k \lambda}(t) e^{-i\delta_{31}t},
	\end{equation}
\begin{equation}
	\label{eq:subeq2}
	{\dot{A}_{2}(t)}  = -\sum_{k \lambda} g_{k,\lambda}^{(21)} A_{k \lambda}(t) e^{-i\delta_{21}t},
\end{equation}
\begin{equation}
	\label{eq:subeq3}
	{\dot{A}_{k \lambda}(t)}  = g_{k,\lambda}^{(21)} A_{2}(t) e^{i\delta_{21}t}+g_{k,\lambda}^{(31)} A_{3}(t) e^{i\delta_{31}t},
\end{equation}
\end{subequations}
where $\delta_{ij}=\omega_{k}-\omega_{ij}$  is the detuning of the radiation mode frequency $\omega_k$ from the atomic transition frequency $\omega_{ij}$.  Performing a simple time integration of Eq.~(\ref{eq:subeq3}) and substituting the result into Eqs.~(\ref{eq:subeq1}) and (\ref{eq:subeq2}),  we obtain

\begin{eqnarray}
	\label{eq:8}
	{\dot{A}_{3}(t)}&=&-\int_{0}^{t} G_{33}(t-t')A_{3} (t') dt' \nonumber \\
	 & &- e^{i\omega_{32}t} \int_{0}^{t} G_{32}(t-t') A_{2} (t') dt',
\end{eqnarray}

\begin{eqnarray}
	\label{eq:9}
	{\dot{A}_{2}(t)}&=&-\int_{0}^{t} G_{22}(t-t')  A_{2} (t') dt' \nonumber \\
& &- e^{-i\omega_{32}t} \int_{0}^{t} G_{23}(t-t') A_{2} (t') dt',
\end{eqnarray}
where \hspace{1mm} $ G_{ij}(t-t')=\sum_{k,\lambda} g^{i1}_{k,\lambda} g^{j1}_{k,\lambda}e^{-i\delta_{j1}(t-t')}, (i,j=2,3)$
are the delay Green's functions. The resulting memory kernel strongly depends on the photon density of states of the field in the reservoir\cite{woldeyohannes1999coherent}. In the following, the solutions of the time-dependent amplitudes for the cases of isotropic photonic band gap and free space are given.

\subsection{An atom in PBG Material}\label{secBell}
Let's consider a three-level atom is embedded in an isotropic photonic crystal\cite{john1994spontaneous,yang2000sontaneous}. The resonant transition frequencies $\omega_{31}$ and $\omega_{21}$ are assumed to be near the band gap, so there exist a strong quantum interference between the  transitions from the upper levels to the lower level\cite{yang2000sontaneous}. 
The dispersion relation of an isotropic photonic crystal near the band edge $\omega_c$  could be approximately expressed by
\begin{equation}
	\label{eq:10}
	\omega_k = \omega_c+A(k-k_0)^2,
\end{equation}
 where $A=\omega_c /k_0^2$  is the curvature near $\omega_c$. This corresponds to a density of states of the form, 
 $ \rho(\omega)\propto \Theta(\omega-\omega_c)(\omega-\omega_c)^{-1/2}$
and the Heaviside step function $\Theta(\omega-\omega_c)$ characterizing the cut-off behavior\cite{entezar2010controllable}.  

Using the isotropic dispersion relation and assuming that the dipole moments are parallel to each other and $g^{21}_{k,\lambda}=g^{31}_{k,\lambda}=g_{k,\lambda}$ for simplicity, we arrive at the following solutions for $A_3(t)$ and $A_2(t)$ \cite{yang2000sontaneous}
\begin{equation}
	\label{eq:11}
	A_{3}(t) = \sum_{j}\frac{f_{1}(x^1_{j})}{Z'(x_{j}^1)}e^{i x_{j}^1 t} + \sum_{j}\frac{f_{2}(x^2_{j})}{H'(x_{j}^2)}e^{i x_{j}^2 t}-R_{3}(t),
\end{equation}
\

\begin{equation}
	\label{eq:12}
	A_{2}(t) = e^{-i \omega_{12}t}\left[ \sum_{j}\frac{f_{3}(x^1_{j})}{Z'(x_{j}^1)}e^{i x_{j}^1 t} + \sum_{j}\frac{f_{4}(x^2_{j})}{H'(x_{j}^2)}e^{i x_{j}^2 t}\right]-R_{2}(t).
\end{equation}
Refer to Appendix A for more details.

\subsection{An atom in Free Space}\label{secBell}

In this subsection, we study the case when the three-level atom is in free space with the photon dispersion relation $\omega_k =ck$\cite{woldeyohannes1999coherent}. The free space case will be helpful to compare and interpret the results of the PBG case, which will be discussed in the following sections. For such a photon dispersion relation, the Green's function is given by  

\begin{equation}
	\label{eq:29}
	G_{ij}(t-t')=\sqrt{\gamma_{i1}\gamma_{j1}}\delta(t-t'),
\end{equation}
whit $\gamma_{ij}=\omega_{ij}^3 d_{ij}^2/6\pi \epsilon_{0} \hslash c^3, (i,j=2,3)$. By substituting Eq.~(\ref{eq:29}) into Eq.~(\ref{eq:8}) and Eq.~(\ref{eq:9}), the time dependence of the amplitudes can be obtained as\cite{woldeyohannes2003coherent}

\begin{equation}
	\label{eq:30}
	A_{3}(t)=e^{-\gamma_{31}t}\sum_{j=1}^{2}C_{j}e^{q_{j}t},
\end{equation}

\begin{equation}
	\label{eq:31}
	A_{2}(t)=e^{-(\gamma_{31}+i\omega_{32})t}\sum_{j=1}^{2}B_{j}e^{q_{j}t},
\end{equation}
where,
\begin{equation}
	\label{eq:32}
	q_{1,2}=\frac{\lambda}{2} \pm \sqrt{\left(\frac{\lambda}{2}\right)^2+{(\bar{\gamma})}^2},
\end{equation}

\begin{equation}
	\label{eq:33}
	\lambda=\gamma_{31}-\gamma_{21}+i\omega_{32},\hspace{0.5cm} \bar{\gamma}=\sqrt{\gamma_{31}\gamma_{21}},
\end{equation}

\begin{equation}
	\label{eq:34}
	C_{j}=\frac{q_{k}\cos (\frac{\theta}{2})+\bar{\gamma} e^{i\phi}\sin(\frac{\theta}{2})}{q_{k}-q_{j}},\hspace{0.5cm}(k=1,2;k\neq j),
\end{equation}

\begin{equation}
	\label{eq:35}
	B_{j}=-q_{j}C_{j}/ \bar{\gamma}.
\end{equation}

By using the Eq.~(\ref{eq:6}), the density operator of the atom-field system is given by 
$\rho_{af}=\ket{\psi}\bra{\psi}$, and the reduced density
matrix of the atom can be obtained as
\begin{equation}
	\label{eq:36}
	\rho_a(t)=Tr_f\{ {\rho_{af}(t)}\}=
	\begin{pmatrix}
		\rho_{33} & \rho_{32} & 0\\
		\rho_{23} & \rho_{22}  & 0\\
		0 & 0& \rho_{11}
	\end{pmatrix},
\end{equation}
where
\begin{align}
	\label{eq:37}
	\begin{split}
		\rho_{33} &= \left | A_{3}(t) \right | ^2, 
		\rho_{22} = \left | A_{2}(t) \right | ^2, 
		\rho_{11} = 1-\rho_{33}-\rho_{22}, \\
		\rho_{32} &=\rho^{*}_{23}=A_{3}(t)A_{2}^*(t).\\
		\end{split}
\end{align}
In the following, we study the dynamics of the quantum properties of the three-level atom in PBG compared to free space.

\section{Quantum Fisher Information}\label{secBell}

Multiparameter quantum estimation theory is a branch of quantum metrology that deals with the estimation of multiple parameters in quantum systems. It provides a framework for understanding and quantifying the precision with which these parameters can be measured\cite{szczykulska2016multi}. In this section, we are interested in studying the dynamic variation in the estimation precision of parameters $\theta$ and $\phi$ encoded into the initial state of the three-level atom. We want to understand how the initial values of the relative phase $\phi$ and the different relative positions of the upper levels from the forbidden gap affect the sensitivity in the measurement of these parameters. To this end, let us briefly review the notion of multiparameter quantum estimation theory.

We use the quantum Fisher information matrix (QFIM) to investigate the optimal precision for estimating the parameters $\theta$ and $\phi$ which is given by\cite{liu2020quantum}

\begin{equation}
	\label{eq:38}
	F(\theta,\phi)=
	\begin{pmatrix}
		\mathcal{F}_{\theta \theta}(t) & \mathcal{F}_{\theta\phi}(t)\\
		\mathcal{F}_{\phi\theta}(t) & \mathcal{F}_{\phi\phi}(t)  \\
	\end{pmatrix},
\end{equation}
where the coefficients are defined as
\begin{equation}
	\label{eq:39}
	\mathcal{F}_{ij}:= Tr(\rho(t)L_i L_j+\rho(t)L_j L_i),
\end{equation}
with $i,j\in {\theta ,\phi}$, and $L_i (L_j )$ is the symmetric logarithmic derivative (SLD),  which is determined by the equation 
\begin{equation}
	\label{eq:40}
	\partial_i \rho(t)= \frac{1}{2}(\rho(t) L_i+L_i  \rho(t)).
\end{equation}
Since the SLD operator is a Hermitian operator, the QFIM is Hermitian. The QFIM is indeed a crucial tool for understanding the fundamental limits of precision in parameters estimation in quantum systems. It can be used to derive the quantum Cramér-Rao bound, which provides a lower bound on the achievable variance of any unbiased estimator. It satisfies the following inequality
\begin{equation}
	\label{eq:41}
	\Sigma(\theta, \phi) \geq F^{-1}(\theta,\phi)
\end{equation}
where $\Sigma$ is the covariance matrix for the parameters $\theta$ and $\phi$, and $F^{-1}(\theta,\phi)$ is the inverse matrix of the QFIM $F(\theta,\phi)$. The corresponding quantum  Cramér-Rao bounds for independent estimations of the parameters $\phi$ and $\theta$ can be obtained as \cite{Metro22,Metro23}
\begin{equation}
	\label{eq:42}
	\Delta\phi\geq \frac{1}{\sqrt{F_{\phi}}}, \hspace{5mm}	\Delta\theta\geq \frac{1}{\sqrt{F_{\theta}}}
\end{equation}
where $\Delta \theta$ and $\Delta\phi$ represent the minimum achievable uncertainties in the estimation of $\phi$ and $\theta$, respectively. 
Based on Eq.~(\ref{eq:39}), the quantities $F_{\theta}=\mathcal{F}_{\theta\theta}$ and $F_{\phi}=\mathcal{F}_{\phi\phi}$ are the diagonal elements of the QFIM, computed as $F_{\theta}=Tr(\rho(t) {L_\theta}^2)$ and $F_{\phi}=Tr(\rho(t) {L_\phi}^2)$. As Eq.~(\ref{eq:42}) shows, in order to achieve better measurement precision, it is indeed desirable to maximize the QFI during the time evolution. 

 \begin{figure}[t!]
 	\includegraphics[width=0.5\textwidth]{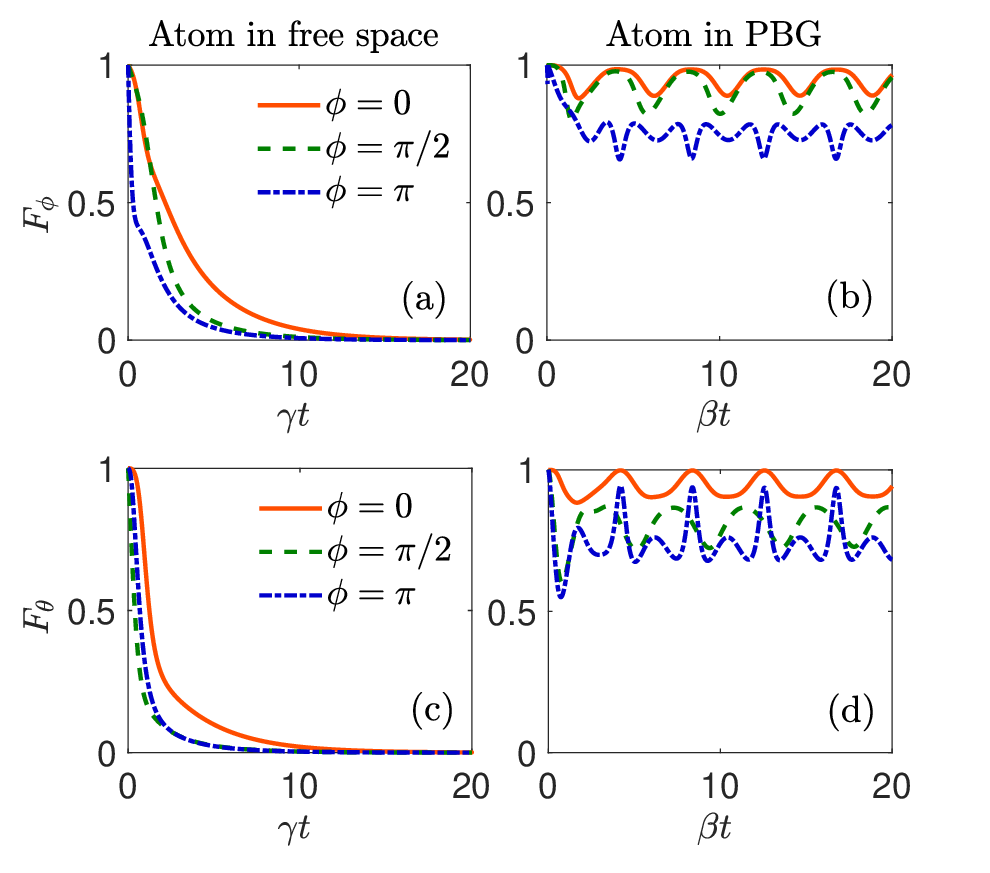}
 	\caption{Dynamical behavior of quantum Fisher information for different initial relative phase values $\phi$, with  $\theta=\pi /2$. The left and right panels correspond to the atom being located in free space and the PBG with $\omega_{3c}=-1 \beta$, respectively.}
 	\label{Fig.2}
 \end{figure}

\begin{figure}[t!]
	\includegraphics[width=0.5\textwidth]{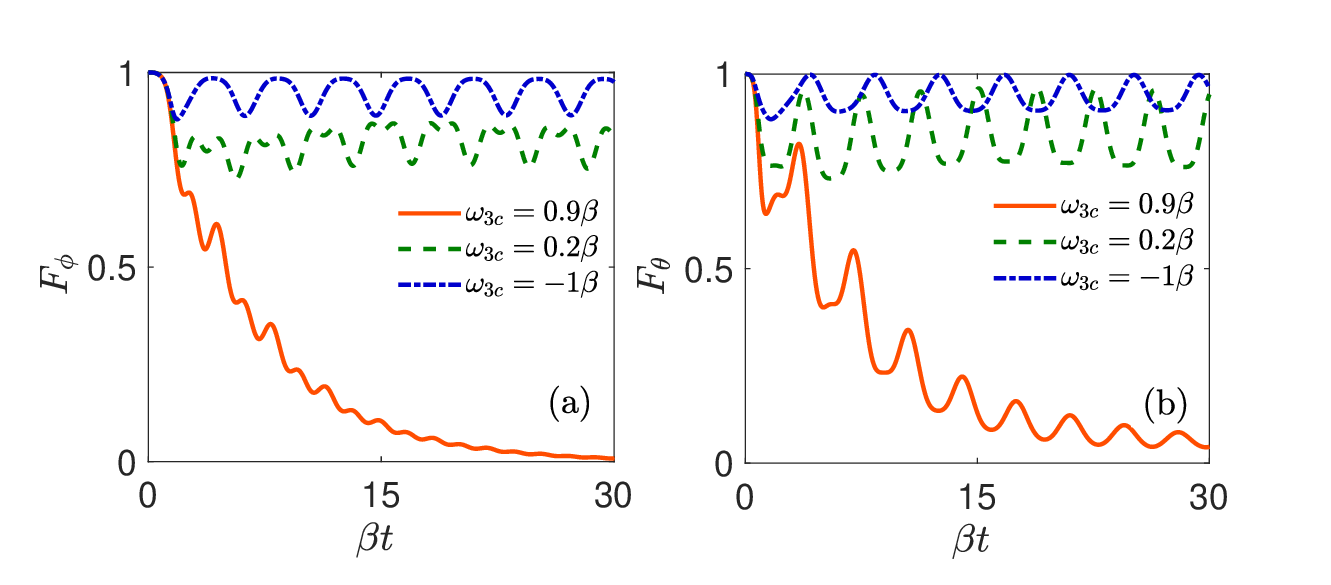}
	\caption{Dynamical behavior of quantum Fisher information for different values of $\omega_{3c}$ with $\theta=\pi/2$ and $\phi=0$. }
	\label{Fig.3}
\end{figure}

Figure \ref{Fig.2} shows the dynamical behavior of $F_{\phi}$ and $F_{\theta}$ as a function of the scaled time for different initial relative phase values $\phi$, with  $\theta=\pi /2$. The left and right panels correspond to the atom being located in free space and the photonic band gap with $\omega_{3c}=-1 \beta$, respectively. As expected, both QFIs quickly approach zero in free space, regardless of the initial relative phase value (see Figs. \ref{Fig.2}(a) and \ref{Fig.2}(c)). In contrast,  $F_{\phi}$ and $F_{\theta}$ exhibit different behavior when the atom is located in the photonic band gap. Despite variations in the behavior of the QFIs across different initial relative phase values, all of them display a monotonic oscillatory pattern within a range close to 1 and can be effectively protected (see Figs. \ref{Fig.2}(b) and \ref{Fig.2}(d)). It is worth noting that setting the initial relative phase to $\phi=0$ leads to the maximum value of the QFIs during the time evolution within the photonic band gap. Therefore, the selection of an appropriate relative phase difference for the initial states significantly influences the quantum Fisher information.

Figure \ref{Fig.3} depicts the dynamical behavior of  $F_{\phi}$ and $F_{\theta}$ as a function of the scaled time for the different relative positions of the upper levels from the forbidden gap, with $\phi=0$ and  $\theta=\pi /2$. When the different relative position of the upper level is $\omega_{3c}=0.9 \beta$,  the QFIs exhibit damped oscillations and eventually decay to zero as time passes. This implies that the QFIs cannot be preserved in this case. 
As the relative position of the upper level from the forbidden gap decreases to $\omega_{3c}=0.2 \beta$ and $\omega_{3c}=-1 \beta$, the process of losing information slows down, and more photons are absorbed back into the quantum system (qutrit). As a result, spontaneous emission is mainly suppressed, and the quantum Fisher information (QFI) exhibits periodic oscillatory behavior with constant amplitude. When the atomic transition frequency is well inside the band gap with $\omega_{3c}=-1 \beta$, the QFIs are closer to their maximum value of 1, compared to the case of $\omega_{3c}=0.2 \beta$. So one can find that in a strongly non-Markovian environment, the decay of the QFI can be suppressed, which means that the information about the qutrit can be preserved for a longer time. 

 \begin{figure}[t!]
	\includegraphics[width=0.47\textwidth]{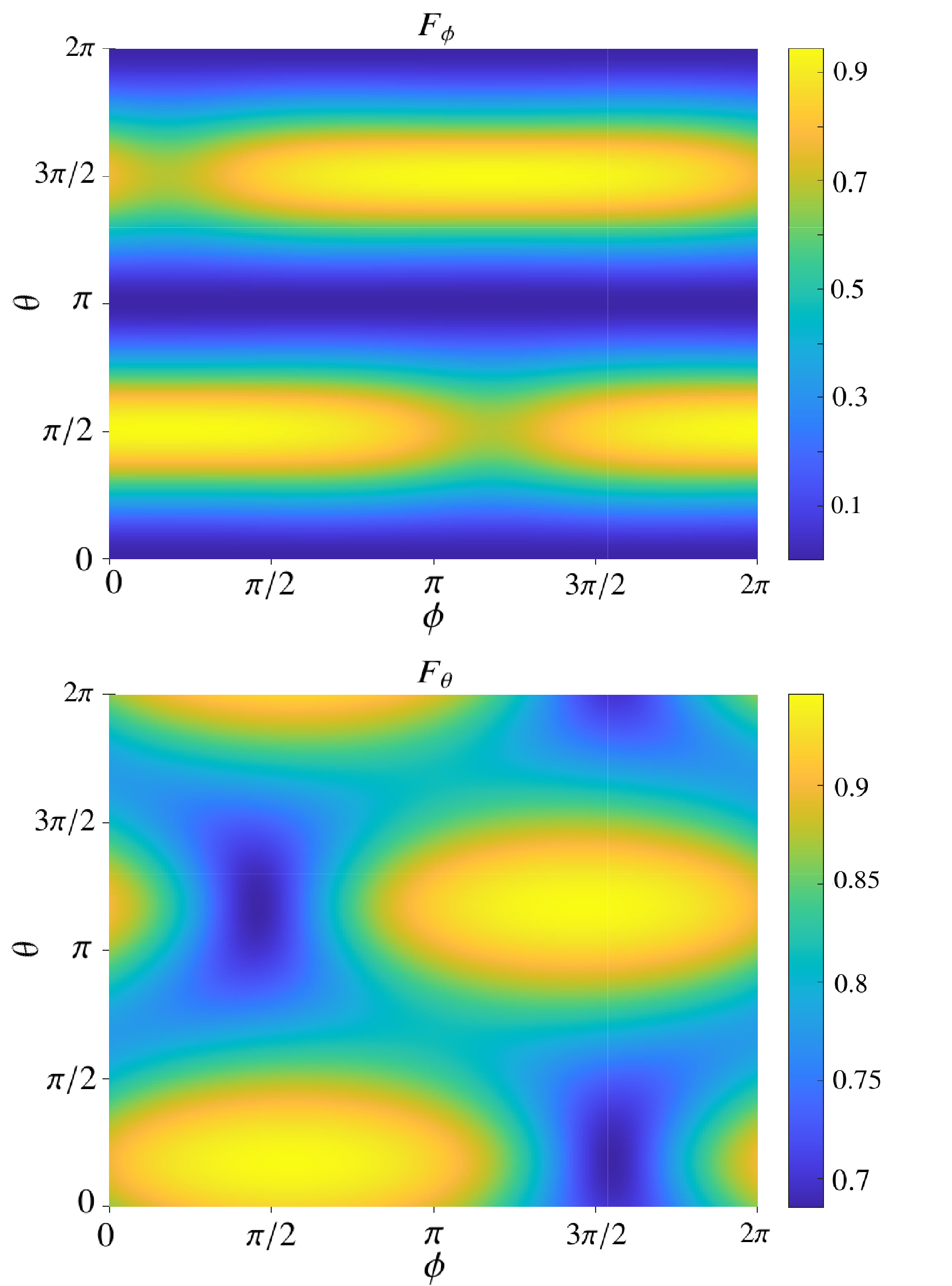}
	\caption{Density plot of $F_\phi$ and $F_\theta$ as a function of $\phi$ and $\theta$ in photonic band gap with $\omega_{3c}=-1 \beta$. }
	\label{Fig.4}
\end{figure}

Based on the previous plots, we expect that changing the value of  $\theta$ while keeping $\phi$ constant has a symmetric effect on the evolution of QFIs. To gain a broader perspective on this characteristic, in Fig. \ref{Fig.4}, we present a contour plot that displays the values of $F_\phi$ and $F_\theta$ over a wide range of initial state parameters $\phi$ and $\theta$ with $\omega_{3c}=-1 \beta$. By choosing appropriate initial state parameters $\theta$ and $\phi$, we can obtain a high value for $F_{\phi}$ and $F_{\theta}$.

\begin{figure}[t!]
	\includegraphics[width=0.5\textwidth]{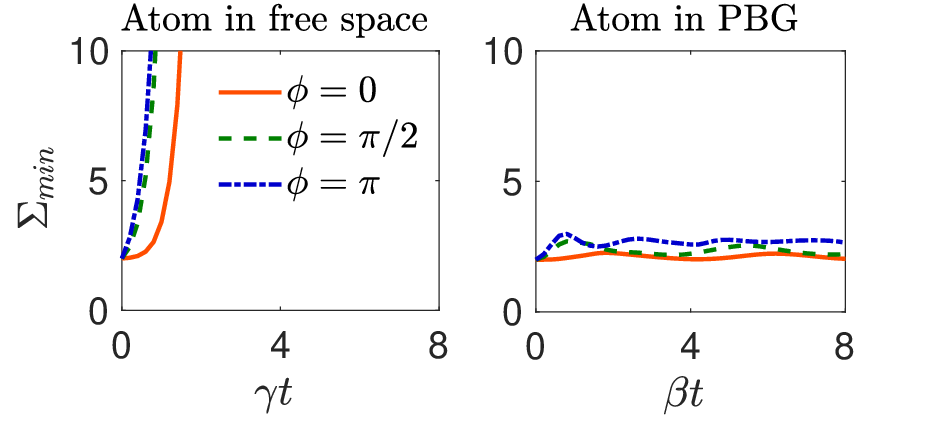}
	\caption{Dynamical behavior of $\Sigma_{min}$ for different valuse of $\phi$ with $\theta=\pi/2$. The left and right panels correspond to the atom being located in free space and the PBG with $\omega_{3c}=-1 \beta$, respectively. }
	\label{Fig.5}
\end{figure}

\begin{figure}[t!]
	\includegraphics[width=0.4\textwidth]{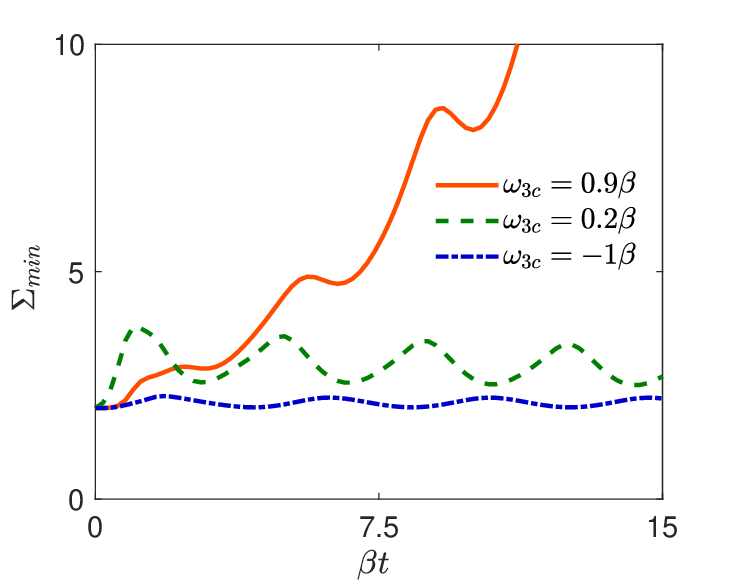}
	\caption{Dynamical behavior of $\Sigma_{min}$ for different valuse of $\omega_{3c}$  with $\theta=\pi/2$ and $\phi=0$.  }
	\label{Fig.6}
\end{figure}

In the following, we used the QFIM approach to calculate the QCRB for the simultaneous estimation of both parameters $\phi$ and $\theta$. Figure 5 illustrates the dynamical behavior of  $\Sigma_{min}=\min{(\Sigma(\theta,\phi))}$ as a function of scaled time for different initial relative phase values $\phi$, with $\theta=\pi/2$. The left and right panels correspond to the atom being located in free space and the photonic band gap with $\omega_{3c}=-1 \beta$, respectively.

In free space,  $\Sigma_{min}$ increases dramatically for all initial relative phases $\phi$ (see Fig. \ref{Fig.5}(a)). This could be due to decoherence, and the absence of any enhancement mechanisms to preserve the information. In contrast, in the photonic band gap, $\Sigma_{min}$ exhibits a relatively stable behavior for all initial relative phases $\phi$ (see Fig. \ref{Fig.5}(b)). The oscillatory behavior around an initial constant value with a small amplitude indicates that  $\Sigma_{min}$ is relatively well-preserved over time, so there is a potential for measuring both parameters simultaneously in the photonic band gap. It is evident that the optimal two-parameter estimation can be obtained for $\phi=0$.

Figure \ref{Fig.6} shows the time evolution of  $\Sigma_{min}$ as a function of scaled time for the different relative positions of the upper levels from the forbidden gap, with $\phi=0$ and  $\theta=\pi /2$.
As anticipated, when the relative position $\omega_{3c}=0.9 \beta$, $\Sigma_{min}$ rapidly increases over time. Furthermore, $\Sigma_{min}$ exhibits an oscillatory pattern with a consistent amplitude for both $\omega_{3c}=0.2 \beta$ and $\omega_{3c}=-1 \beta$. However, the amplitude of oscillation is quite small when $\omega_{3c}=-1 \beta$. Overall, for attaining the highest level of optimal sensitivity in the two-parameter measurement, the photonic band gap material with $\omega_{3c}=-1 \beta$, is suitable. 

\section{Quantum Coherence}\label{secBell}

One of the fundamental features that separate quantum physics from classical physics is the idea of quantum superposition, which leads to quantum entanglement and quantum coherence. Quantum coherence is one of the most important signatures of quantum theory and can be considered as a necessary condition for quantum correlation\cite{yao2015quantum, streltsov2015measuring, radhakrishnan2016distribution}. From the application point of view, it is a key resource for quantum information processing\cite{pan2017complementarity, mortezapour2018coherence} and quantum metrology\cite{giovannetti2011advances}.  We now investigate, the effect of the relative phase of the initial state and the different relative positions of the upper levels from the forbidden gap on the time evolution of the quantum coherence in our three-level atom.

Many measures have been proposed for quantum coherence [1]. Among them we focus here on the $ l_1 $ norm of coherence $C_{l_1}$. For an arbitrary quantum state $\rho$, $ C_{l_1} $ is quantified by\cite{baumgratz2014quantifying}

\begin{figure}[t!]
	\includegraphics[width=0.5\textwidth]{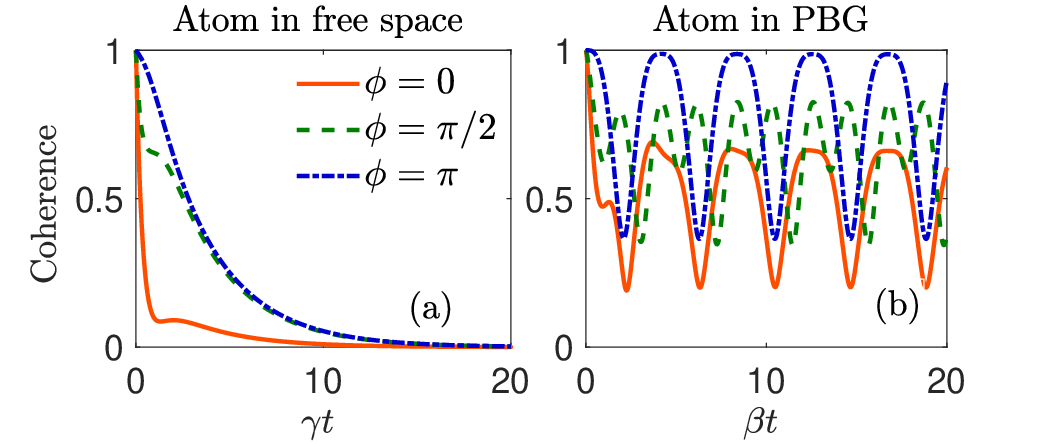}
	\caption{Dynamical behavior  of quantum coherence for different values of $\phi$ with $\theta=\pi/2$. The left and right panels correspond to the atom being located in free space and the PBG with $\omega_{3c}=-1 \beta$, respectively.}
	\label{Fig.7}
\end{figure}

\begin{equation}
	\label{eq:43}
	C_{l_{1}}(\rho(t))=\sum_{i\neq j} |\rho_{ij}(t)|
\end{equation}
where $\rho_{ij} (i\neq j)$ are the off-diagonal elements of the density matrix $\rho(t)$.

\begin{figure}[t!]
	
	\includegraphics[width=0.4\textwidth]{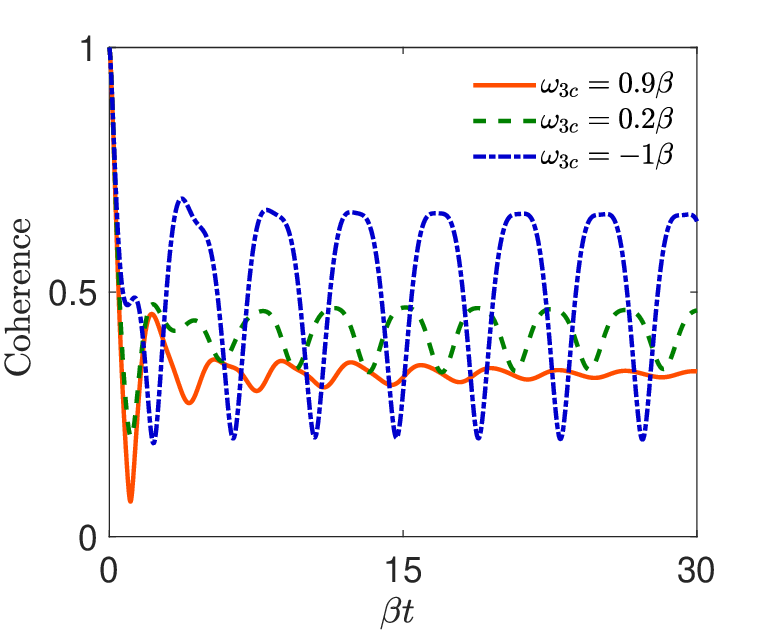}
	\caption{Dynamical behavior of quantum coherence for different values of $\omega_{3c}$ with $\theta=\pi/2$ and $\phi=0$. }
	\label{Fig.8}
\end{figure}

In Fig. \ref{Fig.7}, we plot the dynamical behaviors of the quantum coherence versus scaled time for different values of the initial relative phase $\phi$, with $\theta=\pi /2$. Figures \ref{Fig.7}(a) and \ref{Fig.7}(b), respectively, correspond to the case when the atom is located in free space and photonic band gap material with $\omega_{3c}=-1 \beta$. The initial relative phase $\phi$ strongly affects the dynamical behavior of quantum coherence in both cases.
Figure \ref{Fig.7}(a) reveals that although in free space (Markovian regime) the quantum coherence initially strongly depends on the relative phase of the initial states, it tends monotonously to zero in a short time for all values of the relative phase $\phi$. Losing the information from an open quantum system to the environment causes the quantum coherence to generally decrease over time.
It is noticed that for the case where the atom is in photonic band gap with $\omega_{3c}= -1\beta$, the quantum coherence has a regular oscillation behavior for all values of relative phase $\phi$. The amplitudes of the periodic oscillations do not decrease during the time evolution. In addition, when $\phi = \pi$ the quantum coherence oscillates with no decaying (see Fig. \ref{Fig.7}(b)).

Figure \ref{Fig.8} depicts the time evolution of the quantum coherence for the different relative positions of the upper levels from the forbidden gap with $\phi=0$ and $\theta = \pi/2$. It is obvious that when the relative position of the level $\ket{a_3}$ from the forbidden gap is $\omega_{3c} = 0.9\beta$, the oscillation amplitude of the quantum coherence decays faster and reaches its steady-state value. Moreover, when the relative position of the upper level from the forbidden gap decreases to $\omega_{3c} = 0.2\beta$ or $\omega_{3c} = -1\beta$, the quantum coherence displays an oscillatory behavior with a constant amplitude.

Physically, the oscillatory behavior of the quantum coherence in photonic band gap material is interpreted as memory effects of the non-Markovian environment.
In addition, by choosing the appropriate relative phase of the initial state and the position of the photon band gap, the most optimal state can be obtained that maintains the highest amount of quantum coherence in the system.

\section{Non-Markovianity}\label{secBell}
In this section, we are interested in studying the influences of system-environment interaction on the dynamics of the open system. Usually the interaction between an open quantum system and its environment results in information and coherence loss. However, it can be observed that in a photonic band gap, the non-Markovian effect can sustain the quantum Fisher information and quantum coherence, preventing their rapid loss. These memory effects can lead to richer dynamics, characterized by oscillations, revivals, or even information backflow.
In this paper, the Hilbert-Schmidt speed measure (HSS) \cite{Jahromi,Mahdavipour} is utilized to quantify non-Markovianity, which refers to the departure of a quantum system's evolution from a Markovian or memoryless process.

\begin{equation}
	\label{eq:44}
	HSS(\rho(\phi))=\sqrt{\frac{1}{2}Tr\left[\left(\frac{d\rho(\phi)}{d\phi}\right)^2\right]}
\end{equation}

It quantifies how much information about the future dynamics of the system can be gained by observing the environment. HSS, as an efficient tool in quantum metrology, helps in understanding the flow of information between a system and its environment.
Generally for an initial state as

\begin{equation}
	\label{eq:45}
	\ket{\psi}= \frac{1}{\sqrt{n}}(e^{i\phi}\ket{\psi_{1}}+...+\ket{\psi_{n}})
\end{equation}
(here $\phi$ is an unknown phase shift and \{$\ket{\psi_{i}}\}$ with $ i=1,...,n$ is a complete and orthonormal basis of the Hilbert space) the HSS witness of non-Markovianity is obtained by

\begin{equation}
	\label{eq:46}
	\chi(t)=\frac{dHSS(\rho(\phi))}{dt}>0
\end{equation}
in which $\rho(\phi)$ denotes the evolved state of the system.

 \begin{figure}[t!]
	\includegraphics[width=0.5\textwidth]{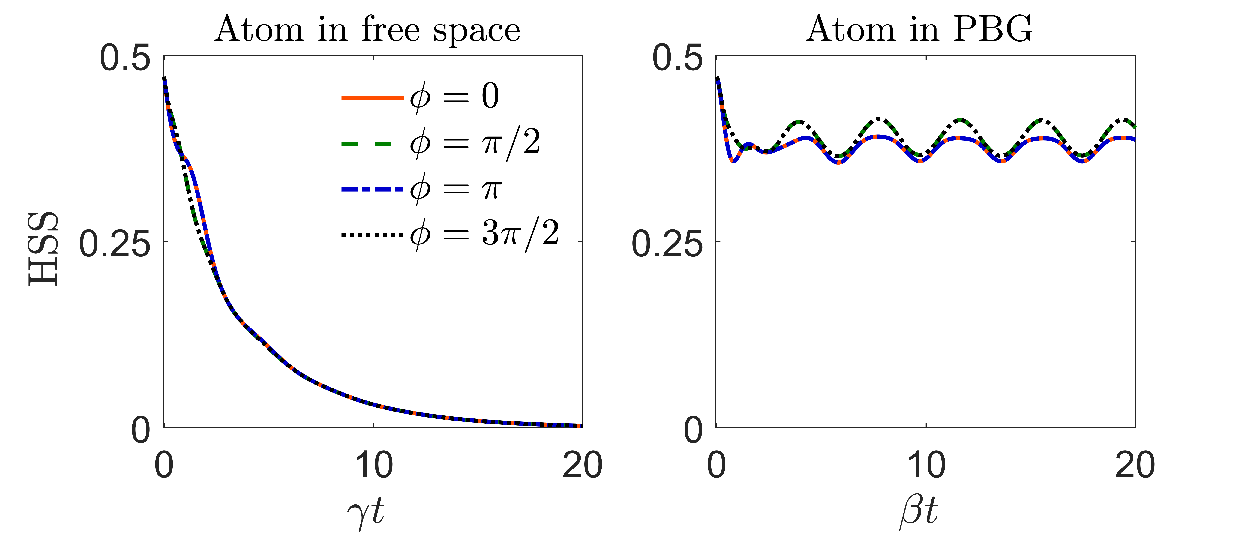}
	\caption{Dynamical behavior of HSS for different values of $\phi$. The left and right panels correspond to the atom being located in free space and the PBG with $\omega_{3c}=-1 \beta$, respectively.}
	\label{Fig.9}
\end{figure}

To properly study non-Markovianityby by using the HSS dynamics for the V-type three-level atom, the qutrit must be initially prepared in a state as described by Eq. (29), which is given by:
\begin{equation}
	\label{eq:47}
	\ket{\psi}= \frac{1}{\sqrt{3}}(\ket{a_3}+e^{i\phi}\ket{a_2}+\ket{a_1})
\end{equation}
The elements of the reduced density matrix for the atom, in this initial state can be derived as follows:

\begin{align}
	\label{eq:48}
	\begin{split}
		\rho_{33} &= \left | A_{3}(t) \right | ^2, 
		\rho_{22} = \left | A_{2}(t) \right | ^2, 
		\rho_{11} = 1-\rho_{33}-\rho_{22}, \\
		\rho_{32} &=\rho^{*}_{23}=A_{3}(t)A_{2}^*(t),
		\rho_{31} = \rho^{*}_{23}=\frac{A_{3}(t)}{\sqrt{3}} ,\\
		\rho_{21} &= \rho^*_{12}=\frac{A_{2}(t)}{\sqrt{3}}.\\
	\end{split}
\end{align}
Here,  $A_{3}(t)$ and $A_{2}(t)$ can be found by substituting $\cos (\frac{\theta}{2})=\sin (\frac{\theta}{2})=\frac{1}{\sqrt{3}}$ into Eqs. (11) and (12) for the atom in PBG crystal, and into from Eqs. (15) and (16) for the atom in free space.

Figure \ref{Fig.9} displays the dynamical behavior of HSS as a function of the scaled time for different initial relative phase values $\phi$. The left and right panels correspond to the atom being located in free space and photonic band gap with $\omega_{3c}=-1 \beta$, respectively. It is observed that both situation show the same dynamical behavior of the HSS for the pairs $\phi=0,\pi$ and $\phi=\pi /2,3\pi /2 $. As is evident in the case of free space, HSS exhibits Markovian behavior for all values of $\phi$ and shows no significant changes in its time derivative  $\chi(t)$ (see Fig. \ref{Fig.9}(a)). This means that in free space, information always flows to the environment. On the other hand, in the case of PBG, the time evolution of HSS exhibits fluctuations for all values of $\phi$. The dynamics in this case are highly non-Markovian, and the time derivative of  $\chi(t)$ intermittently becomes positive (see Fig. \ref{Fig.9}(b)). This indicates that during these times, the Hilbert-Schmidt speed increases and leads to the backflow of information from the environment to the system.

\begin{figure}[t!]
	\includegraphics[width=0.4\textwidth]{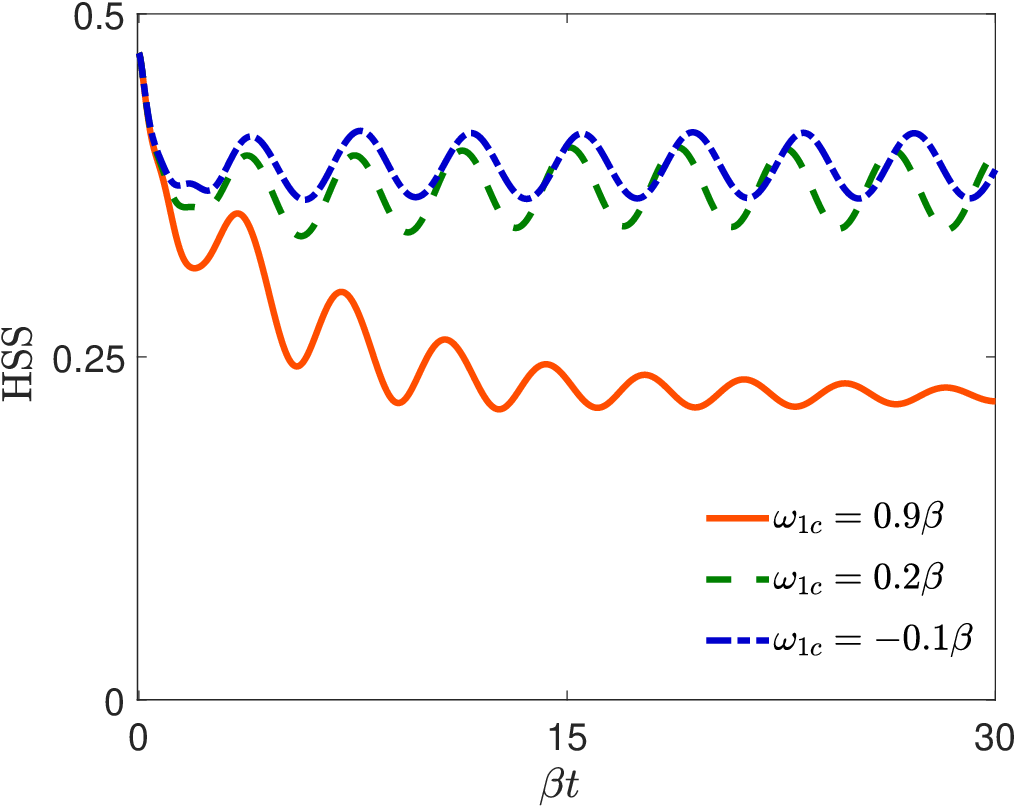}
	\caption{Dynamical behavior of HSS for different values of $\omega_{3c}$ with $\phi=0$.}
	\label{Fig.10}
\end{figure}

The dynamical behavior of HSS as a function of scaled time is illustrated in Fig. \ref{Fig.10} for different relative positions of the upper levels from the forbidden gap, with $\phi=0$. It is evident that when the relative position of the upper level from the forbidden gap decreases to $\omega_{3c} = 0.2\beta$ or $\omega_{3c} = -1\beta$,  the dynamics of HSS exhibit oscillatory patterns. These oscillations indicate that the system's interaction with its environment has led to feedback or memory effects, resulting in the flow of information to the environment and backflow to the system. When the relative position of the upper level is $\omega_{3c} = 0.9\beta$,  the HSS shows displays damped oscillations, gradually approaching a steady-state over time. This implies the flow of information to the environment and, occasionally, the backflow of information from the environment to the system. Ultimately, over an extended duration, the information flows completely into the environment. This analysis shows that structured reservoirs, such as photonic band gap materials, enhance the non-Markovianity of the system's dynamics.
\section*{CONCLUSION}
We have analyzed the dynamical behavior of quantum Fisher information, quantum coherence, and non-Markovianity of a V-type three-level atom embedded in free space or a photonic band gap crystal while considering all spontaneous emissions. Our results imply that the photonic band gap crystal, as a structured environment, significantly influences the preservation and enhancement of these quantum features. Moreover, we have examined in detail the effect of the initial relative phase values encoded in the atomic state and the relative positions of the upper levels within the forbidden gap on the evolution of quantum Fisher information, quantum coherence, and non-Markovianity.
It is shown that when the atomic transition frequency is well inside the band gap with $\omega_{3c}=-1 \beta$  and $\phi=0$ , $F_{\phi}$ and $F_{\theta}$ are closer to their maximum value of 1. This indicates that in a strongly non-Markovian environment, the decay of the quantum Fisher information can be suppressed, allowing for longer preservation of information about the qutrit. Additionally, we have observed that the optimal two-parameter estimation can be obtained for $\phi=0$ in the photonic band gap with $\omega_{3c}=-1 \beta$.

It is noticed that for the case where the atom is in a photonic band gap with $\omega_{3c}= -1\beta$, the quantum coherence exhibits an oscillating behavior for all values of the relative phase $\phi$, corresponding to non-Markovian evolution. The amplitudes of the periodic oscillations do not decrease during the time evolution.
The dynamical behavior of non-Markovianity measured by HSS has confirmed our analysis of the dynamics of quantum coherence.

\appendix
\section*{APPENDIX}
\renewcommand{\theequation}{A.\arabic{equation}}
\setcounter{equation}{0}  
Using the isotropic dispersion relation  of Eq.~ (\ref{eq:10}) we can evaluate the corresponding Green's function as\cite{woldeyohannes2003coherent}  

\begin{equation}
	\label{eq:A1}
	G_{ij}(t-t') = \beta^{3/2}_{j} \frac{e^{i[\delta_{j1}(t-t')-\pi/4]}}{\sqrt{\pi(t-t')}} ; (i,j=2,3),
\end{equation}
where $\beta_{j}^{3/2}=[(\omega_{j} d_{j})^2/6\pi \epsilon_{0}\hslash](k_{0}^3/ \omega_{c}^{3/2})$ and $ (j=2,3)$. By taking the Laplace transform of Eqs.~(\ref{eq:8}) and (\ref{eq:9}), we obtain.
{\footnotesize 
	\begin{equation}
		\begin{aligned}
		\label{eq:A2}
		A_{3}(s&)=\\ &\frac{\cos (\frac{\theta}{2}) (s-i \omega_{32} + G_{22}(s-i\omega_{32})) - e^{i\phi}\sin(\frac{\theta}{2}) G_{32}(s-i\omega_{32})}{[s+G_{33}(s)][(s-i\omega_{32})+G_{22}(s-i\omega_{32})]-G_{23}(s)G_{32}(s-i\omega_{32})},\\
	\end{aligned}
	\end{equation}

\begin{equation}
	\begin{aligned}
		A_{2}(s&-i\omega_{32})=  \\
		& \frac{e^{i\phi}\sin(\frac{\theta}{2})[s+G_{33}(s)]-\cos(\frac{\theta}{2})G_{23}(s)}{[s+G_{33}(s)][(s-i\omega_{32})+G_{22}(s-i\omega_{32})]-G_{23}(s)G_{32}(s-i\omega_{32})}, \\
	\end{aligned}
\end{equation}
}
here G(s) is the Laplas transform of the Green's function in Eq.~(\ref{eq:A1}), which can be derived in the following form,

\begin{equation}
	\label{eq:A4}
	G_{33}(s)=-\frac{\beta_{3}^{3/2}}{ \sqrt{is+\omega_{3c}}},
\end{equation}

\begin{equation}
	\label{eq:A5}
	G_{22}(s-i\omega_{32})=-\frac{\beta_{2}^{3/2}}{ \sqrt{is+\omega_{3c}}},
\end{equation}

\begin{equation}
	\label{eq:A6}
	G_{32}(s-i\omega_{32})=G_{23}(s)=-\frac{\beta_{3}^{3/4} \beta_{2}^{3/4}}{ \sqrt{is+\omega_{3c}}},
\end{equation}
where  $\omega_{3c}=\omega_{3}-\omega_{c}$. In the following discussion, we consider that the two dipole moments are parallel to each other and for the sake of  simplicity $g^{21}_{k,\lambda}=g^{31}_{k,\lambda}=g_{k,\lambda}$ so, $G_{33}(s)=G_{22}(s-i\omega_{32})=G_{23}(s)=G(s)$ and $\beta_3=\beta_{2}= \beta$. In this case, by using the inverse Laplace transform the amplitudes $A_3(t)$ and $A_2(t)$ can be obtained as \cite{yang2000sontaneous}.

\begin{equation}
	\label{eq:A7}
	A_{3}(t) = \sum_{j}\frac{f_{1}(x^1_{j})}{Z'(x_{j}^1)}e^{i x_{j}^1 t} + \sum_{j}\frac{f_{2}(x^2_{j})}{H'(x_{j}^2)}e^{i x_{j}^2 t}-R_{3}(t),
\end{equation}
\\

\begin{equation}
	\label{eq:A8}
	A_{2}(t) = e^{-i \omega_{12}t}\left[ \sum_{j}\frac{f_{3}(x^1_{j})}{Z'(x_{j}^1)}e^{i x_{j}^1 t} + \sum_{j}\frac{f_{4}(x^2_{j})}{H'(x_{j}^2)}e^{i x_{j}^2 t}\right]-R_{2}(t),
\end{equation}
where
\begin{equation}
	\label{eq:A9}
	R_{3}(t) = \frac{\beta^{3/2}}{\pi\sqrt{i}}\int_{0}^{\infty}\frac{e^{-bt}\sqrt{x}a[\cos (\frac{\theta}{2})a+e^{i\phi}\sin(\frac{\theta}{2})b]}{xb^2 a^2 -i[2x-i(\omega_{3c}+\omega_{2c})]^2\beta^3}dx,
\end{equation}

\begin{equation}
	\label{eq:A10}
	R_{2}(t) = \frac{\beta^{3/2}}{\pi\sqrt{i}}\int_{0}^{\infty}\frac{ e^{-at}\sqrt{x} b [\cos (\frac{\theta}{2}) a +e^{i\phi}\sin(\frac{\theta}{2})b]}{x b^2 a^2 -i[2x-i(\omega_{3c}+\omega_{2c})]^2\beta^3}dx.
\end{equation}
with $a=x-i\omega_{2c}$ , $b=x-i\omega_{3c}$ and the functions are defined as follows

\begin{equation}
	\label{eq:A11}
	f_{1}(x)=\cos (\frac{\theta}{2})(x-i\omega_{32})-\frac{[\cos (\frac{\theta}{2})-e^{i\phi}\sin(\frac{\theta}{2})]\beta^{3/2}}{\sqrt{ix+\omega_{3c}}},
\end{equation}

\begin{equation}
	\label{eq:A12}
	f_{2}(x)=\cos (\frac{\theta}{2})(x-i\omega_{32})+\frac{[\cos (\frac{\theta}{2})-e^{i\phi}\sin(\frac{\theta}{2})]\beta^{3/2}}{\sqrt{ix+\omega_{3c}}},
\end{equation}

\begin{equation}
	\label{eq:A13}
	f_{3}(x)=e^{i\phi}\sin(\frac{\theta}{2})x-\frac{[e^{i\phi}\sin(\frac{\theta}{2})-\cos (\frac{\theta}{2})]\beta^{3/2}}{\sqrt{ix+\omega_{3c}}},
\end{equation}

\begin{equation}
	\label{eq:A14}
	f_{4}(x)=e^{i\phi}\sin(\frac{\theta}{2})x+\frac{[e^{i\phi}\sin(\frac{\theta}{2})-\cos (\frac{\theta}{2})]\beta^{3/2}}{\sqrt{ix+\omega_{3c}}},
\end{equation}

\begin{equation}
	\label{eq:A15}
	Z(x)=x(x-i\omega_{32})-\frac{[2x-i\omega_{32}]\beta^{3/2}}{\sqrt{ix+\omega_{3c}}},
\end{equation}

\begin{equation}
	\label{eq:A16}
	H(x)=x(x-i\omega_{32})+\frac{[2x-i\omega_{32}]\beta^{3/2}}{\sqrt{ix+\omega_{3c}}},
\end{equation}
where $	Z'(x)=\frac{dZ(x)}{dx}$ and  $H'(x)=\frac{dH(x)}{dx}$. In Eqs.~(\ref{eq:A7}) and (\ref{eq:A8}), $x_{j}^1$ are the root of $Z(x)=0$ in the region $Im(x_{j}^1)>\omega_{3c}$ or $Re(x_{j}^1)>0$; $x_{j}^2$ are the root of $H(x)=0$ in the region $Im(x_{j}^2)<\omega_{3c}$ and $Re(x_{j}^2)<0$. 
\section*{ACKNOWLEDGMENTS}
The authors acknowledge S.Nafise Mousavi for the valuable assistance in the discussions.
\bibliography{ref}

 \end{document}